\documentclass[11pt,a4paper]{article}
\usepackage[latin1]{inputenc}
\usepackage{jcappub}
\usepackage{amsmath}
\usepackage{amsfonts}
\usepackage{amssymb}
\usepackage{graphicx}
\usepackage{epsfig}
\usepackage{subfigure}
\usepackage{color}
\usepackage{amssymb}

\def\beq{\begin{equation}}
\def\eeq{\end{equation}}
\def\baq{\begin{eqnarray}}
\def\eaq{\end{eqnarray}}

\newcommand{\bk}{{\bf k}}

\providecommand{\f}[2]{\frac{{#1}}{{#2}}}

\title{Dark matter from gravitational particle production at reheating}

\author[a]{Tommi Markkanen}
\author[b]{and  Sami Nurmi}

\affiliation[a]{Department of Physics, King's College London, Strand, London WC2R 2LS, UK}
\affiliation[b]{Department of Physics, University of Jyv\"{a}skyl\"{a},  P.O. Box 35, FI-40014 University of Jyv\"{a}skyl\"{a}, Finland}

\abstract{We show that curvature induced particle production at reheating generates adiabatic dark matter if there are non-minimally coupled spectator scalars weakly coupled to visible matter.  The observed dark matter abundance implies an upper bound on spectator masses $m$ and non-minimal coupling values $\xi$. For example, assuming 
quadratic inflation, instant reheating and a single spectator scalar with only gravitational couplings,  the observed dark matter abundance is obtained for $m\sim 0.1$ GeV and $\xi \sim 1$. Larger mass and coupling values of the spectator are excluded as they would lead to overproduction of dark matter. }

\emailAdd{tommi.markkanen@kcl.ac.uk}
\emailAdd{sami.t.nurmi@jyu.fi}

\begin{document}

\maketitle
\section{Introduction}

Cosmological observations are consistent with inflation driven by a single scalar field \cite{Ade:2015lrj,Ade:2015xua}. Extensions of the Standard Model (SM) of particle physics however generically contain several scalars.  This suggests that there could be many dynamically irrelevant spectator scalars present during inflation. The Higgs field observed at the LHC is an example of such a spectator unless its high energy potential would significantly deviate from the SM prediction \cite{Espinosa:2007qp}.  While the spectators have no impacts on inflationary dynamics, they may have significant observational ramifications ranging from dark matter to primordial perturbations \cite{McDonald:1993ex,Enqvist:2001zp,Lyth:2001nq,Moroi:2001ct,Linde:1996gt,Mollerach:1989hu,Dvali:2003em,Kofman:2003nx,Enqvist:2013kaa,Enqvist:2014zqa,Nurmi:2015ema,Enqvist:2015sua}.  

For an interacting scalar field, a non-minimal coupling to spacetime curvature $\xi R \phi^2$ is 
always generated by the renormalization group flow in curved space. 
While the coupling is irrelevant in the weak gravitational fields of the current universe, it can play a crucial role at the very high energies of the early universe and especially during inflation. 
It has been proposed that the SM Higgs could act as the inflaton provided its curvature coupling is very large $\xi \gg 1$ \cite{Bezrukov:2007ep}. The stability of the SM vacuum in the early universe also crucially depends on the non-minimal coupling  \cite{Espinosa:2007qp,vacstab2,vacstab1,vacstab3,Fairbairn:2014zia,Enqvist:2014bua,Hook:2014uia,Bezrukov:2014ipa,Espinosa:2015qea,Herranen:2014cua, Herranen:2015ima,Kamada:2014ufa,Moss:2015gua}. For the measured Higgs and top quark masses the SM vacuum is only metastable but with a lifetime significantly longer than the age of the universe \cite{deg,but}. However, the Higgs fluctuations generated during inflation could easily have triggered a transition to the true vacuum.  The fact that this did not happen requires either $\xi\gtrsim 0.1$ \cite{Herranen:2014cua}, a low inflationary scale $H\lesssim 10^{11} $ GeV or new physics beyond the Standard Model. In \cite{Herranen:2015ima} it was further shown that the production of Higgs particles at reheating through the curvature induced time-dependent mass may also destabilise the vacuum for $\xi\gtrsim 1$.  


In this work we show that non-minimally coupled spectator scalars with weak couplings to visible matter constitute a generic dark matter component. During reheating the inflaton oscillations cause the curvature scalar $R$ to oscillate between positive and negative values. Negative values of $R$ generate tachyonic mass terms $m_{\rm eff}^2\sim \xi R$ for the non-minimally coupled scalars which leads to explosive particle production. The mechanism is  generic for $\xi \gtrsim 1$ when the curvature term gives a large contribution to effective masses of the spectator fields. Dynamics of the instability is similar to the tachyonic preheating scenario \cite{Bassett:1997az,Tsujikawa:1999jh,Dufaux:2006ee}. A similar effect was also recently discussed in the context of SM vacuum stability \cite{Herranen:2015ima}. 

If the spectators are weakly coupled to visible matter, the particles generated at reheating will not decay and therefore form a dark matter component. If there were no isocurvature perturbations, each Hubble patch undergoes exactly the same reheating dynamics with shifted time coordinates. The amount of spectator particles is the same in each patch and the dark matter generated through the curvature coupling is therefore adiabatic. In this sense the setup bears some similarity to the freeze-in scenario \cite{Hall:2009bx} which also leads to adiabatic dark matter although the dark matter particles were never thermalised with the visible matter. 

In this work we focus on the simplest case with a single spectator scalar with the non-minimal coupling $\xi R \chi^2$ and no non-gravitational interactions. We consider the parameter regime $\xi\gtrsim 1$ , where $\chi$ is effectively massive during inflation and will not form a primordial condensate \cite{Enqvist:2013kaa,Nurmi:2015ema,Figueroa:2015rqa,Enqvist:2015sua}. In this simple setup we show that the tachyonic production of $\chi$ particles could easily account for the observed dark matter abundance $\Omega_{\rm DM}h^2\simeq 0.12$. For example, for quadratic inflation and instant reheating, the correct dark matter abundance is obtained for singlet mass $m\sim 10$ GeV provided that $\xi \sim 1$. Larger masses are excluded as they would lead to overproduction of dark matter. For larger values of $\xi$ the mass bound gets even tighter. The quantitative constraints are alleviated if the inflationary scale is lower or the reheating is not instantaneous but qualitatively the picture remains the same.

Our analysis is based on the approximation where the matter fields are quantized and the metric is viewed as classical. This  is expected to be a valid approximation for energetically subdominant spectator fields when spacetime dynamics are dominated by the inflaton decoupled from spectators. This means that our calculation can be approached in the framework of quantum field theory on a curved background \cite{Birrell:1982ix} (for a recent application, see \cite{Czerwinska:2015xwa}). In contrast to other recent studies dealing with a non-minimally coupled theory \cite{George:2013iia,George:2015nza}, our calculation is done in the Jordan frame, in which our analysis is simpler as there the model can be solved non-perturbatively without any issues related to renormalization. Related work on inflation and reheating with non-minimally coupled scalar fields can be found in \cite{Watanabe:2015eia,Kaiser:2013sna,Schutz:2013fua,DeCross:2015uza}, and in particular the generation of isocurvature via non-minimal couplings is discussed in \cite{Schutz:2013fua}.
 
Our sign choices are (+,+,+) in the classification of \cite{Misner:1974qy}.

\section{The setup}

To present the idea, we concentrate on the simplest possible case. We consider a non-minimally coupled scalar singlet $\chi$ with no non-gravitational interactions 
\beq
S = \int d^4 x \sqrt{-g}\left(- \frac{1}{2}\nabla^{\mu}\chi\nabla_{\mu}\chi - \frac{1}{2}m^2\chi^2 -\frac{\xi}{2} \chi^2 R\right)\ .
\eeq
Inclusion of self-interactions for $\chi$ would not affect our general conclusions. However, the assumption of very weak couplings to visible matter is crucial for the generated singlet particles to survive as stable dark matter. For example, a coupling to SM Higgs $\lambda_{\chi h} \chi^2 h^2$ allowed by symmetries must be in the range $\lambda_{\chi h}< 10^{-7}$ so that the $\chi$ particles can never thermalise with the SM fields, see e.g. \cite{Hall:2009bx}.  Throughout the paper we assume $\chi$ is an energetically subdominant spectator field $\rho_{\chi} \ll 3 H^2 M_{\rm pl}^2$. 

During inflation the curvature scalar is given by $R= 12 H_{}^2$, where $H_{}= \dot{a}/a$ is the Hubble rate. We concentrate on the parameter regime $\xi\gtrsim 1$, where the field $\chi$ is effectively massive during inflation $m^2_{\rm eff} = m^2 + 12 \xi H_{}^2 \gtrsim H_{}^2$ and does not get displaced from the vacuum. 

We assume the inflationary dynamics can be parameterised by a single scalar field $\phi$, the inflaton. The curvature perturbation $\zeta$ is then completely determined by the inflaton fluctuations and  is conserved on superhorizon scales.  After the end of inflation the inflaton starts to oscillate and eventually decays into relativistic degrees of freedom heating up the universe. We assume that the inflaton potential during reheating can be modelled by the quadratic form
\beq
\label{Vreh}
V(\phi) =\frac{1}{2}m_{\phi}^2 \phi^2\ , \qquad \phi<\phi_{\rm osc}~. 
\eeq
Here $\phi_{\rm osc}$ denotes the field value at the peak of the first inflaton oscillation which we hereafter refer to as the onset of oscillations. We further assume that the inflaton energy density dominates the universe such that the Hubble rate at the onset of oscillations reads  
\beq
\label{H0}
H_{\rm osc} = \frac{m_{\phi}\phi_{\rm osc}}{M_{\rm pl}\sqrt{6}} ~. 
\eeq

Note that apart from eq. (\ref{Vreh})  we do not specify the form of the inflaton potential. In particular, the potential may differ from the quadratic form in the inflationary regime $\phi>\phi_{\rm osc}$.  Consequently, in deriving our general results, we treat the mass scale $m_{\phi}$ and the amplitude $\phi_{\rm osc}$ in Eq. (\ref{Vreh}) as free parameters. Their relation to the measured CMB perturbations is determined only after the complete inflaton potential is specified. 
 
After the onset of inflaton oscillations the curvature scalar, given by the trace of Einstein equation $G_{\mu\nu}M_{\rm pl}^{2}=T_{\mu\nu}$, reads
\beq
\label{Rosc}
R=M_{\rm pl}^{-2} \left(4 V(\phi)-\dot{\phi}^2\right)\ .
\eeq
Therefore, $R$ becomes an oscillatory function which changes sign each time the inflaton field enters the regime $\dot{\phi}^2 > 4V(\phi)$. The average energy density of the oscillating inflaton corresponds to non-relativistic matter and consequently the average value of $R$ is given by 
$\langle R\rangle =3H^2\propto a^{-3}$. As the inflaton eventually decays, the universe becomes radiation dominated and the scalar curvature vanishes $R=0$ at the classical level.

\section{Curvature induced particle production at reheating}

The oscillatory curvature scalar induces a time-dependent mass term $m^2_{\rm eff}= m^2 + \xi R$ for the singlet $\chi$.  We assume its bare mass $m$ is small compared to the curvature induced part at the beginning of oscillations $m^2\ll \xi H_{\rm osc}^2$. The square of the effective mass $m^2_{\rm eff} \simeq \xi R$ then becomes negative around zero crossings of the oscillating inflaton when $R$ is negative. This leads to tachyonic generation of $\chi$ particles \cite{Bassett:1997az}. 

The equation of motion for $\chi$ can be written in terms of the rescaled field 
$\tilde{\chi} = \chi/a$, and conformal time $ds^2=a^2(-d\eta^2+d\mathbf{x}^2)$ in Fourier space as
\beq
\tilde{\chi}''_{\bk}+ \bigg[\bk^2 + a^2 m^2+ a^2 \left(\xi-\frac{1}{6}\right) R \bigg] \tilde{\chi}_{\bk} = 0 \  .\label{eom3}
\eeq
For $m^2\ll \xi H_{\rm osc}^2$, the effective mass goes negative for $\dot{\phi}^2 > 4V(\phi)$ according to eq. (\ref{Rosc}). In this regime there are tachyonic modes for which  
\beq
\label{tachyonic}
\omega_{\bk}^2= \bk^2 + a^2 m^2 + a^2 \left(\xi-\frac{1}{6}\right) R  < 0 \, .
\eeq

These modes are exponentially amplified $\chi_{k} \propto e^{|\omega_k|\eta}$ which may result in very efficient particle production, even after a single inflaton oscillation \cite{Herranen:2015ima}.  This is very similar to preheating from the tachyonic instability analysed in detail in \cite{Dufaux:2006ee}. The instability may last over several inflaton oscillations but since $\langle R\rangle\propto a^{-3}$ the total yield of $\chi$ particles is dominated by the first few oscillation cycles. Here we use a conservative estimate for the number of $\chi$ particles and account only for their generation over the first inflaton oscillation. Including particle production from later times could only tighten the bounds we find below.

The solution for the oscillating inflaton field reads $\phi(t) \simeq \Phi(t){\rm cos}(mt)$ where $\Phi\propto a^{-3/2}$ is a slowly varying amplitude. Following \cite{Dufaux:2006ee,Herranen:2015ima}, we approximate the inflaton amplitude by a constant over the first oscillation and set $\Phi = \phi_{\rm osc}$.  Eq. (\ref{eom3}) for the $\chi$ field can then be recast  in the form of a Mathieu equation \cite{Dufaux:2006ee}
\beq
{\f{d^2{(a^{3/2}\chi}_{\mathbf{k}})}{dz^2}+\bigg[A_k-2q\cos(2z)\bigg]a^{3/2}\chi_{\mathbf{k}}=0,\qquad z=m_\phi t\,,\label{eq:mathieu}}
\eeq
\beq{\nonumber A_k= \f{\mathbf{k}^2}{a^2m_\phi^2}+\xi \f{\phi_{\rm osc}^2}{2 M_{\rm pl}^2},\qquad q=\f{3\phi_{\rm osc}^2}{4 M_{\rm pl}^2}\bigg(\f{1}{4}-\xi\bigg)~.}
\eeq
Here we have neglected the bare $\chi$ mass which we assumed to be small $m^2\ll \xi H_{\rm osc}^2$. The condition for a mode to become tachyonically amplified, eq. (\ref{tachyonic}), when expressed in the Mathieu form then reads 
\beq 
-A_k+2q\cos(2z)\equiv\Omega_{\mathbf{k}}^2>0\,.\label{eq:tach}
\eeq
As shown in \cite{Dufaux:2006ee}, the occupation number $n_\mathbf{k}$ of tachyonically excited modes after the first inflaton oscillation is given by 
\beq
{n_\mathbf{k}=e^{2X_\mathbf{k}}\,,\qquad X_\mathbf{k}=\int_{\Delta z}\Omega_\mathbf{k}\,dz 
~, 
\label{eq:occapp1}} 
\eeq
where $\Delta z$ denotes the region of $z$ values during the first oscillation for which eq. (\ref{eq:tach}) is satisfied. 

Following \cite{Herranen:2015ima} we can derive a lower bound of the produced particles by including only the superhorizon modes, $k < a_{\rm osc}H_{\rm osc}$. Using eqs. (\ref{eq:mathieu}), (\ref{eq:tach}) and (\ref{eq:occapp1}) we get 
\begin{align}
X_\mathbf{k}&=\int_{\Delta z}\bigg\{-\f{\mathbf{k}^2}{a^2m_\phi^2}-\xi \f{\phi_{\rm osc}^2}{2 M_{\rm pl}^2}+2\bigg[\f{3\phi_{\rm osc}^2}{4 M_{\rm pl}^2}\bigg(\f{1}{4}-\xi\bigg)\bigg]\cos(2z)\bigg\}^{1/2}\,dz\nonumber \\ &\geq\f{\phi_{\rm osc}}{M_{\rm pl}}\sqrt{\f{\xi}{2}}\int_{\Delta z}\bigg\{-1-\f{1}{3\xi}-3\bigg(1-\f{1}{4\xi}\bigg)\cos(2z)\bigg\}^{1/2}\,dz\,,\label{eq:x}
\end{align}
where we have also used eq. (\ref{H0}) for $H_{\rm osc}$. For $\xi \gg 1$ we can discard the $\xi^{-1}$ terms in the wavy brackets of (\ref{eq:x}), which after performing the integral results in the occupation number of $\chi$ particles on superhorizon scales as
\beq n_\mathbf{k}\approx\exp\left\{\sqrt{\xi}\f{2\phi_{\rm osc}}{M_{\rm pl}}\right\}\, \label{eq:occapp}\, ,
\eeq
in agreement with \cite{Dufaux:2006ee, Herranen:2015ima}. For $\xi\gg 1$ this clearly is an exponentially large quantity provided that $\phi_{\rm osc}\sim M_{\rm pl}$.
The variance of $\chi$ particles after the first inflaton oscillation then reads\footnote{Note that in deriving this result we have treated the inflaton amplitude $\phi_{\rm osc}$ as a constant over the first oscillation and hence neglected a non-exponential contribution $\langle\chi^2\rangle \sim H^2$, i.e. standard particle production due to a smooth spacetime expansion \cite{Zeldovich:1971mw}. This is justified in the limit $\sqrt{\xi}\phi_{\rm osc}/M_{\rm pl}\gg 1$ where the tachyonic production dominates, in the opposite case eq. (\ref{chivariance}) should be regarded as a rough order of magnitude estimate which however suffices for our current purposes.}   
\beq \langle \chi^2 \rangle_{\rm osc}\approx \int^{a_{\rm osc}H_{\rm osc}}_{0} \f{d\vert\mathbf{k}\vert\, \mathbf{k}^2n_\mathbf{k}}{2\pi^2a_{\rm osc}^3\sqrt{\xi R_{\rm osc}}} 
\sim\left(\f{H_{\rm osc}}{2\pi}\right)^2\f{2\exp\left\{\sqrt{\xi}\f{2\phi_{\rm osc}}{M_{\rm pl}}\right\}}{3\sqrt{3\xi}}\, .
\label{chivariance}\eeq 

Note that the amplification of superhorizon modes is a consequence of the coherently oscillating inflaton background which is nearly the same in each causal patch, hence there is no violation of causality \cite{Bassett:1999mt,Finelli:1998bu}. From (\ref{chivariance}) we see that the spectrum of the generated $\chi$ particles scales as $P_{\chi}\propto k^3$ and peaks around the horizon scale.

We reiterate that generation of the large variance (\ref{chivariance}) for a non-minimally coupled scalar $\chi$ relies on two conditions. First, the bare mass of the $\chi$ particle must be small at the onset of inflaton oscillations, $m^2\ll \xi H_{\rm osc}^2$, so that the tachyonic amplification may take place. Second, our analysis assumes the $\chi$ field is energetically subdominant during reheating $\rho_{\chi} \ll 3 H^2 M_{\rm pl}^2$. If $\xi$ is large enough, this condition could be violated by the tachyonic production of $\chi$ particles. However, in this case $\rho_{\chi}\sim 3 H^2 M_{\rm pl}^2$ which leads to overproduction of cold dark matter if the $\chi$ particles become non-relativistic before the matter radiation equality which always happens for $m\gtrsim 10^{-3}$ eV,  see Section~4. Therefore, for $m \gtrsim 10^{-3}$ eV the tachyonic generation of $\chi$ particles can never violate the energy condition $\rho_{\chi} \ll 3 H^2 M_{\rm pl}^2$ in the observationally viable regime. 

The results given above were obtained assuming the $\chi$ field has no non-gravitational interactions. Adding a self-interaction term ${\lambda}\chi^4$ to the model would not change the results provided that $\lambda\langle\chi^2\rangle \ll \xi R$ during the tachyonic phase. Using the case of quadratic inflation as an example, $H_{\rm osc} \sim 10^{13}$GeV and $\phi_{\rm osc}\sim 0.3 M_{\rm pl}$,  we then find from eq. (\ref{chivariance}) that for $\xi \lesssim 10^{-3}$ the self-interactions can be neglected for $\lambda\ll 10^{-2}$. Larger self-couplings would not necessarily remove the tachyonic growth but the instability bands and growth rates could significantly deviate from the results given above, see \cite{Enqvist:2016mqj} for a related discussion. Similarly, an eventual direct coupling to the inflaton ${g^2}\phi^2\chi^2$ would not affect the tachyonic amplification if $g^2 \phi^2\ll \xi R$, which for $H_{\rm osc} \sim 10^{13}$GeV and $\phi_{\rm osc}\sim 0.3 M_{\rm pl}$ gives $g^2\ll \xi \cdot10^{-10}$.  As the inflaton eventually decays into visible matter, additional model-dependent conditions could follow from our requirement that $\chi$ particles should not thermalise with the visible matter.  

\section{Dark matter abundance}

Assuming that non-gravitational interactions of the $\chi$ field can be neglected, the particles generated by the tachyonic instability are stable and constitute a dark matter component. The amount of particles generated in each Hubble patch is determined by the local value $\phi_{\rm osc}$ of the inflaton field at the onset of oscillations according to eq.  (\ref{chivariance}). Assuming there were no isocurvature perturbations present at 
the end of inflation, $\phi_{\rm osc}$ has a uniform value and hence the number of $\chi$ particles generated in each patch is the same.  Indeed, for adiabatic perturbations each Hubble patch undergoes exactly the same expansion history but with shifted time coordinates. Even though $\chi$ was never in thermal equilibrium with the visible matter the generated dark matter component is adiabatic. 

Neglecting the production of $\chi$ particles after the first inflaton oscillation and accounting only for superhorizon modes we find from eqs. (\ref{eq:occapp}) and (\ref{chivariance}) a lower limit for their number density 
\beq
\label{nchi}
n_{\chi} \simeq \left(\frac{a_{\rm osc}}{a}\right)^3 \frac{H_{\rm osc}^3}{6\pi^2}\exp\left\{\sqrt{\xi}\f{2\phi_{\rm osc}}{M_{\rm pl}}\right\}~.
\eeq
The tachyonic instability follows from the dominance of the negative mass squared contribution in the dispersion relation  (\ref{tachyonic}). Consequently, the excited modes are non-relativistic at the time of their generation. Assuming they remain non-relativistic during the entire reheating stage\footnote{The $\chi$ particles could become relativistic already before the end of reheating if $k/a \sim a_{\rm osc} H_{\rm osc}/a < m_{\rm eff}$ which implies $H_{\rm reh} \lesssim H_{\rm osc} (3\xi)^{-3/2}$. Here we do not consider this possibility.}, their energy density at the end of reheating $t_{\rm reh}$ is given by     
\beq
\label{chienergy}
\rho_{\chi}^{\rm reh} =  m_{\rm eff}(t_{\rm reh})n_{\chi}(t_{\rm reh}) \simeq   \left(\frac{a_{\rm osc}}{a_{\rm reh}}\right)^{9/2} \frac{\sqrt{3\xi}H_{\rm osc}^4}{6\pi^2}\exp\left\{\sqrt{\xi}\f{2\phi_{\rm osc}}{M_{\rm pl}}\right\}~.
\eeq
Here we have used that $m_{\rm eff} = \sqrt{3\xi} H_{\rm osc}(a_{\rm osc}/a)^{3/2} $ during the effectively matter dominated period of inflaton oscillations. 

With (\ref{chienergy}) we can check our previous assertion that the produced energy-density is not gravitationally significant such that back reaction can safely be ignored. The largest possible contribution is obtained with $a_{\rm osc}=a_{\rm reh}$. For the example of chaotic inflation,  $H_{\rm osc}\sim 10^{13}$GeV and $\phi_{\rm osc}\sim 0.3 M_{\rm pl}$, this gives $\rho_{\chi}/(3M_{\rm pl}^2H^2)<10^{-3}$ for\footnote{In the mass range $m > 10^{-3}$eV where the $\chi$ particles constitute cold dark matter at recombination time, eq. (\ref{DMabundance}) below yields the bound $\xi \lesssim 1600$ to avoid overproduction of dark matter. Even for the maximum value $\xi \sim 1600$ we still get $\rho_{\chi}/(3M_{\rm pl}^2H^2)\lesssim 0.2$.} $\xi<10^3$.  We note that this condition results in the same bound as demanding that the scalar curvature-term in the Einstein-Hilbert action is dominant over the non-minimal coupling to curvature $R M_{\rm pl}^2\gg \f{\xi}{2}R\langle\chi^2\rangle$.

As the reheating is completed, the universe becomes radiation dominated and $R=0$ (up to a small contribution from conformal anomaly  neglected for this discussion). The curvature induced $\chi$ mass vanishes and the particles become relativistic. Their energy density then starts to scale as 
\beq
\label{chienergyrel}
\rho_{\chi} = 
\rho_{\chi}^{\rm reh}\left(\frac{a_{\rm reh}}{a}\right)^4 .
\eeq
Eventually the kinetic energies redshift below the bare mass $m$ and the particles again become non-relativistic. Let us estimate when this happens. 

If there are only gravitational interactions between $\chi$ particles they will effectively not scatter off each other. The momentum space distribution (\ref{eq:occapp}) preserves its shape but may get shifted as the effective mass changes. We may thus rewrite the energy density (\ref{chienergyrel}) in the form 
\beq
\label{rhodist}
\rho_{\chi} = \frac{1}{a^4}\int^{k_{\rm max}}_0 \f{d\vert\mathbf{k}\vert\, \mathbf{k}^3n_\mathbf{k}}{2\pi^2}  = \frac{k_{\rm max}^4}{8 \pi^2 a^4} {\rm exp}\left(\frac{2\sqrt{\xi}\phi_{\rm osc}}{M_{\rm pl}}\right)\ .
\eeq
The maximum wavenumber $k_{\rm max}$ is determined by taking the limit $a= a_{\rm reh}$ and comparing the result to eq.  (\ref{chienergy}) 
\beq
k_{\rm max} \sim \xi^{1/8} a_{\rm osc} H_{\rm osc}\left(\frac{H_{\rm reh}}{H_{\rm osc}}\right)^{1/12}\ .\label{max}
\eeq

We may then estimate that after the end of reheating the $\chi$ particles remain relativistic until $k_{\rm max}/a\sim m$ after which their energy density scales as $\rho_{\chi}\propto a^{-3}$. Using eqs. (\ref{chienergy}), (\ref{chienergyrel}) and  (\ref{max}) we get for the energy density at late times $a > a_{\rm nr}$ the result 
\begin{align}
\label{rhochiasymptotic}
\rho_{\chi} &= \bigg(\f{a_{\rm nr}}{a}\bigg)^3\left(\f{a_{\rm reh}}{a_{\rm nr}}\right)^4 \left(\frac{a_{\rm osc}}{a_{\rm reh}}\right)^{9/2} \frac{\sqrt{3\xi}H_{\rm osc}^4}{6\pi^2}\exp\left\{\sqrt{\xi}\f{2\phi_{\rm osc}}{M_{\rm pl}}\right\}
\\ \nonumber&=\frac{\xi^{3/8}}{\sqrt{3}}\left(\frac{T}{T_{\rm reh}}\right)^3 \left(\frac{H_{\rm osc}}{H_{\rm reh}}\right)^{3/4}\frac{m H_{\rm reh}^3}{2\pi^2}\exp\left\{\sqrt{\xi}\f{2\phi_{\rm osc}}{M_{\rm pl}}\right\}\ ,\qquad T < \frac{T_{\rm reh} m}{H_{\rm reh}  \xi^{1/8}}\left(\frac{H_{\rm reh}}{H_{\rm osc}}\right)^{1/4}~.
\end{align}
From the last condition we see that the $\chi$ particles become non-relativistic above the CMB temperature $T_{\rm CMB} \sim 0.3$ eV provided that $m \gtrsim \xi ^{1/8} (T_{\rm reh}/M_{\rm pl})(H_{\rm osc}/H_{\rm reh})^{1/4}$ eV. The non-observation of primordial gravitational waves constrains the inflationary scale from above $H_{\rm inf}\lesssim 10^{14}$ GeV \cite{Array:2015xqh}. Therefore, using $H_{\rm osc} < H_{\rm inf}$ we find that for $m\gtrsim 10^{-3}$ eV the $\chi$ particles always constitute a cold dark matter component at photon decoupling, assuming the reheating process was relatively fast, $H_{\rm reh}\gtrsim 0.01 H_{\rm osc}$. 


For adiabatic perturbations, $\phi_{\rm osc}, H_{\rm osc}$ and $T_{\rm reh}$ have the same values in each Hubble patch. From eq. (\ref{rhochiasymptotic}) we then 
obtain $\delta\rho_{\chi}/\rho_{\chi} = 3\delta T/T$. Comparing this with perturbations of the radiation component $\delta \rho_{\gamma}/\rho_{\gamma} = 4\delta T/T$ we find 
\beq
\frac{\delta \rho_{\chi}}{\rho_{\chi}} - \frac{3}{4}\frac{\delta \rho_{\gamma}}{\rho_{\gamma}} = 0~, 
\eeq
showing that there are no isocurvature perturbations between the $\chi$ field and radiation. Therefore, the tachyonically produced $\chi$ particles indeed constitute an adiabatic dark matter component, provided there was no isocurvature perturbation present at the end of inflation.

Substituting the present photon temperature $T_{0} = 2.725$ K in eq. (\ref{rhochiasymptotic}) we find the dark matter abundance comprised by the non-relativistic $\chi$ particles today 
\beq
\label{DMabundance}
\frac{\Omega_{\chi}h^2}{0.12} \simeq \xi^{3/8}  \left(\frac{m}{10 \rm GeV}\right) \left(\frac{g_{*,{\rm reh}}}{106.75}\right)^{3/2}\left(\frac{T_{\rm reh}}{10^{15} {\rm GeV}}\right)^3 
\left(\frac{H_{\rm osc}}{H_{\rm reh}}\right)^{3/4} {\rm exp}\left(\frac{2\sqrt{\xi}\phi_{\rm osc}}{M_{\rm pl}}\right)~.
\eeq
This is the main result of this work. Here $g_{*,{\rm reh}}$ denotes the effective number of relativistic degrees of freedom at reheating, $T_{\rm reh}$ is the reheating temperature and $H_{\rm osc}$ and $\phi_{\rm osc}$ denote the Hubble rate and inflaton amplitude at the end of inflation and onset of inflaton oscillations. 

\section{Observational constraints}

For $\Lambda$CDM the best fit value of the dark matter abundance today is \cite{Ade:2015xua} \beq \Omega_{\rm CDM}h^2\simeq 0.12\,.\label{eq:obs}\eeq From eq. (\ref{DMabundance}) we see that gravitationally generated $\chi$ particles may easily constitute a sizeable fraction of the observed dark matter. In fact, their energy density might even exceed the observed dark matter abundance which yields an upper limit for the allowed mass $m$ and non-minimal coupling $\xi$. 

\begin{figure}
\begin{center}
\includegraphics[width=0.48\textwidth,trim={3cm 20.55cm  8cm  2cm },clip]{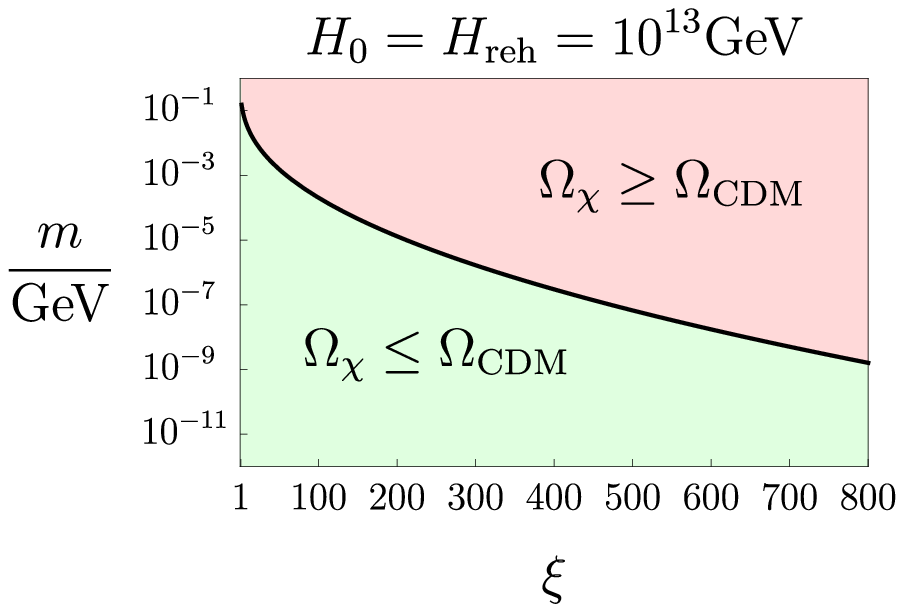}
\includegraphics[width=0.49\textwidth,trim={4.8cm 19.35cm  8cm  5cm },clip]{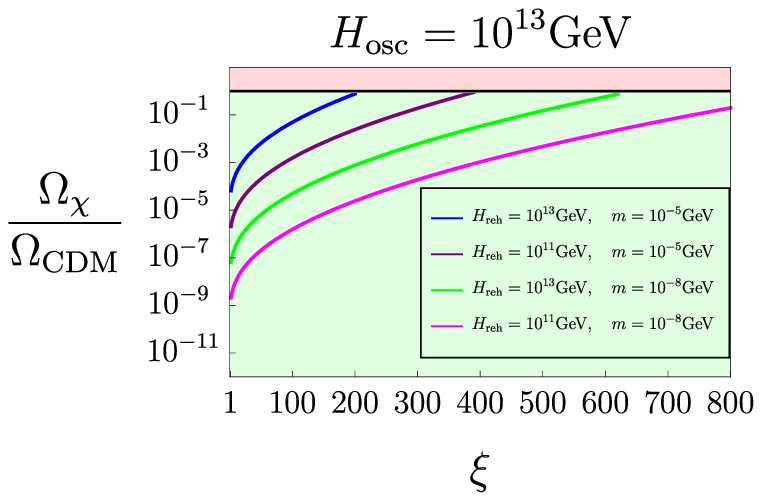}
\caption{\label{fig:regions}In the left panel, we plot the allowed and excluded regions for the mass of the $\chi$ field as a function of the non-minimal coupling $\xi$ for $H_{\rm osc}=H_{\rm reh}=10^{13}$GeV. In the right panel,  we plot $ \Omega_\chi/\Omega_{\rm CDM}$ today as a function of the non-minimal coupling varying the masses and the scale of reheating as $m=10^{-5}$GeV, $m=10^{-8}$GeV and $H_{\rm reh}=H_{\rm osc}$, $H_{\rm reh}=10^{-2}H_{\rm osc}$, respectively. The red region is excluded by the dark matter abundance (\ref{eq:obs}).
}
\end{center}
\end{figure}

As a concrete example, let us consider chaotic inflation with a quadratic potential $V(\phi)= (1/2)m_\phi^2\phi^2$ both during and after inflation. In this case the Hubble rate at the end of slow roll inflation and onset of inflaton oscillations is $H_{\rm osc}\sim 10^{13}$ GeV. From eq. (\ref{tachyonic}) we then see that tachyonic generation of $\chi$ particles occurs for masses $m\lesssim10^{13}$GeV and again for the inflaton value at the peak of the first oscillation we use $\phi_{\rm osc}\sim 0.3 M_{\rm pl}$ \cite{Kofman:1997yn}. Using these input values in (\ref{DMabundance}) and assuming instant reheating temperature with $T_{\rm reh}=[(g_{*,{\rm reh}}\pi^2/30)^{-1}3H_{\rm reh}^2 M_{\rm pl}^2]^{1/4}$  (with the effective number of degrees of freedom close to the SM value $g_{*,{\rm reh}}= 106.75$)  we obtain the results depicted in Fig. \ref{fig:regions}. 
Clearly,  requiring that the energy density of $\chi$ particles does not exceed the observed dark matter abundance, $\Omega_{\chi}\leqslant \Omega_{\rm DM}$, we find a relatively stringent constraint on the mass $m$ and non-minimal coupling $\xi$.  For $\xi \sim 1$ the $\chi$ mass must be in the range $m\lesssim 10^{-1}$ GeV and the bound gets exponentially tighter as $\xi$ is increased. If reheating is not instantaneous, the bounds get alleviated as seen in the left-hand-side of Fig. \ref{fig:regions}. 

While the results shown in Fig. \ref{fig:regions} hold for the specific example of quadratic inflation, the constraints remain qualitatively similar for more generic setups as well.  In particular, our main result eq. (\ref{DMabundance}) only assumes that the inflaton potential can be approximated by a quadratic form near its minimum during reheating. Apart from this constraint eq. (\ref{DMabundance}) applies for any form of the inflaton potential. Changing the potential effectively amounts to changing the values of $\phi_{\rm osc}$, $H_{\rm osc}$ and $H_{\rm reh}$. As can be seen in eq. (\ref{DMabundance}), lowering the inflationary scale and reheating temperature both decrease the energy density of $\chi$ particles and therefore alleviate the constraints on $m$ and $\xi$. 
\section{Conclusions}
Non-minimal couplings of scalar fields to spacetime curvature can be of crucial importance in the very early universe. For interacting scalars, such couplings are inevitably generated at one loop level in curved space. In particular, the stability of the SM vacuum both during inflation and at reheating crucially depends on the Higgs non-minimal coupling.

In this work we have shown that curvature couplings may also play a key role in dark matter generation. We found that gravitational particle production of non-minimally coupled spectator scalars at reheating may constitute a significant dark matter component. For single field inflation, the produced dark matter is adiabatic even if the  spectator fields never were in thermal equilibrium with visible matter. 

The mechanism is based on the generic feature that the curvature scalar becomes negative around minima of the potential of the oscillating inflaton field. This generates tachyonic mass terms $m^2\sim \xi R$ for non-minimally coupled spectator scalars and leads to explosive particle production. If the produced scalar particles are sufficiently decoupled from visible matter they constitute a dark matter component.  
Moreover, if perturbations at the time of reheating are adiabatic, the reheating physics and the number of gravitationally produced particles is the same in each horizon patch. Therefore,  dark matter produced by the mechanism is adiabatic. 

We concentrated on the simplest possible setup with a single non-minimally coupled spectator scalar $\chi$ with no non-gravitational interactions. In this case we found that the gravitational particle production can easily match the observed abundance of dark matter and even exceed it. The observed abundance $\Omega_{\rm CDM}h^2 \simeq 0.12$  therefore implies an upper bound on the scalar mass $m$ and the non-minimal coupling value $\xi$. 

For example, assuming quadratic inflaton $m_{\phi}^2\phi^2$, with the Hubble scale at the onset of reheating given by $H_{\rm osc} \sim 10^{13}$ GeV, we find that masses in the range $m\gtrsim 0.1$ GeV are excluded by dark matter overproduction for $\xi \gtrsim 1$.  It would be interesting to investigate structure formation constraints on this type of dark matter particles which, in the absence of non-gravitational interactions, will retain the out of equilibrium distribution formed by the tachyonic generation process.

While we have considered the simplest case with only one spectator scalar, the results can be straightforwardly extended to more generic setups with several weakly interacting spectator scalars. Finally, we note that if there are non-adiabatic perturbations at the time of reheating the number of gravitationally produced particles will in general vary over different horizon patches. This would lead to generation of isocurvature dark matter which is heavily constrained by observations.  

\section*{Acknowledgments}
The research leading to these results has received funding from the European Research Council under the European Union's Horizon 2020 program (ERC Grant Agreement no.648680).

\end{document}